\documentclass[a4paper,11pt]{article}
\usepackage{pos}
\usepackage{caption}
\usepackage{subcaption}
\newcommand{\RNum}[1]{\uppercase\expandafter{\romannumeral #1\relax}}

\title{Revisiting Two Decades of GRB Observations: Assessing Missed Very High-Energy Detections and Future Prospects
}

\author*[a]{Halim Ashkar}
\author[b]{Stephen Fegan}
\author[c]{Aurélie Sangaré}

\affiliation[a]{Institute of Space Sciences (IEEC-CSIC), Campus UAB, Torre C5, 2a planta, 08193 Barcelona, Spain}

\affiliation[b]{Laboratoire Leprince-Ringuet, École Polytechnique, CNRS, Institut Polytechnique de Paris, F-91128 Palaiseau, France}

\affiliation[c]{Mathematical Institute, University of Oxford, Andrew Wiles Building, Radcliffe Observatory Quarter, Woodstock Road, Oxford, OX2 6GG, U.K.}

\emailAdd{hashkar@ice.csic.es}

\abstract{Gamma-ray bursts (GRBs) are bright flashes of electromagnetic radiation originating from the core collapse of massive stars or the merger of compact objects. It has long been theorized that GRBs can emit very high-energy (VHE) gamma rays that can reach the TeV level. Although current-generation Imaging Atmospheric Cherenkov Telescopes (IACTs), such as H.E.S.S., have been observing GRBs since 2002, the first detection of GRBs by IACTs occurred only 16 years later, in 2018, raising the question of why no detections were made during these years. We investigate all GRBs detected by the Swift Observatory with redshift measurements over the past two decades. Using the phenomenological relationship between X-ray and gamma rays and taking into consideration extragalactic background light absorption effects and instrument response functions, we search for any missed opportunities for GRBs that could have been detected by the three IACTs: H.E.S.S., MAGIC, and VERITAS, and present the best candidates. We find that the missing detections can be explained by the low rate of detectable GRBs at VHE, which we quantify as < 1 per year. We also find that with the future Cherenkov Telescope Array Observatory (CTAO), this rate can increase to 4 per year.}

\ConferenceLogo{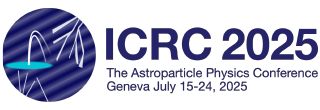}

\FullConference{39th International Cosmic Ray Conference (ICRC2025)\\
 15–24 July 2025\\
Geneva, Switzerland\\}

\begin{document}
\maketitle

\section{Introduction}
Gamma-ray bursts (GRBs) are produced in the core collapse of massive stars or during the merger of compact objects, notably neutron stars. When a new compact object is formed by either of these channels, it accretes the surrounding matter and launches jets of high-energy particles. This jet interacts with the interstellar medium creating shocks and accelerating particles to very high energies. In the magnetic field of the remnant, synchrotron emission takes place which is visible in the radio to X-rays bands. Gamma rays and very high energy (VHE) are produced by the inverse-Compton process in the shocks. 

From 2002 to 2018, Imaging Atmospheric Cherenkov Telescopes (IACTs) followed gamma-ray bursts (GRBs) without recording a single confirmed detection. Aside from the detection of GeV photons by Fermi-LAT, no very-high-energy (VHE) gamma-ray detection was confirmed until 2018. In 2018 and 2019, three confirmed detections were made: GRB 180720B~\cite{GRB180720B} and GRB 190829A~\cite{GRB190829A} by H.E.S.S., and GRB 190114C~\cite{GRB190114C1, GRB190114C2} by MAGIC. This raises several questions: Why were no VHE GRBs detected for 16 years? Did we miss any opportunities to observe GRBs that could have potentially emitted detectable VHE gamma rays? What is the true detection rate of GRBs in the VHE regime for current instruments? What will be this rate with future instruments? 

In the study reported in~\cite{VHEGRB} we retrospectively examine all GRBs detected by the Neil Gehrels Swift Observatory to try and answer these questions. The objective is to simulate the VHE gamma-ray emission from these GRBs as seen by three IACTs, H.E.S.S., MAGIC and VERITAS to answer the questions mentioned previously. Swift GRBs have arcminute to arcsecond localizations. Given that IACTs have wide fields of view, covering these regions is feasible. We identify Swift GRBs that these observatories could have followed based on current observation criteria and delays, aiming to find those potentially detectable in the VHE range based on X-ray flux, distance, observing conditions, and telescope sensitivity—regardless of whether they were actually observed. Section~\ref{sec:2} presents the assumptions used in this study. Section~\ref{sec:3} describes the selection of GRBs. Section~\ref{sec:4} outlines the computation of the expected VHE signal and the treatment of GRB data. Sections~\ref{sec:5} and~\ref{sec:6} present the results.

\section{Assumptions}
\label{sec:2}
We make three key assumptions about VHE emission, inspired by the three detected VHE GRBs.

\begin{itemize}
    \item \textbf{Assumption 1: Relation between X-ray and VHE fluxes.}  
    We assume that the unabsorbed X-ray and VHE energy fluxes are related by:
    \begin{equation}
    \mathrm{\phi_{\gamma}^{(u)} \times F \equiv \phi_{X}^{(u)}}
    \end{equation}
    where where $\mathrm{\phi_{\gamma}^{(u)}}$ and $\mathrm{\phi_{X}^{(u)}}$ are the unabsorbed X-ray energy flux and the intrinsic gamma-ray fluxes respectively and $F$ is a scaling factor derived from observations of the three detected VHE GRBs, typically ranging from 1 to 3. For example:
    \begin{itemize}
        \item GRB\,180720B: $\mathrm{F \sim 1}$ up to 440 GeV at $\sim$10 hours post-burst.
        \item GRB\,190114C: $\mathrm{F \sim 1.5}$ early on, increasing to $\mathrm{F \sim 2.5}$ up to 1 TeV.
        \item GRB\,190829A: $\mathrm{F \sim 3}$ between 0.2 and 4 TeV at $\sim$4 hours post-burst.
    \end{itemize}
    In this work, we adopt $\mathrm{F = 3}$ as observed in GRB\,190829A~\cite{GRB190829A} due to its proximity and minimal extragalactic background light (EBL) absorption.

    \item \textbf{Assumption 2: Temporal decay follows a power-law with matched indices.}  
    We assume both the X-ray and VHE gamma-ray fluxes decay as a power law with the same index $\mathrm{\alpha}$:
\begin{equation}
    \mathrm{\phi_X^{(u)} \propto t^{\alpha_X}, \quad \text{with} \quad \alpha_{\gamma} = \alpha_X}
\end{equation}

    \item \textbf{Assumption 3: VHE photon index.}  
    The VHE photon spectrum is assumed to follow a power-law with a photon index of $\mathrm{\gamma = 2}$.  
    \end{itemize}

\section{GRB selection}
\label{sec:3}
We retrieved all Swift GRB alerts from the GCN\footnote{\url{https://gcn.gsfc.nasa.gov/swift_grbs.html}} network since 2004 and simulated their observability by H.E.S.S., MAGIC, and VERITAS using standardized criteria with darktime and moderate moonlight observations (see~\cite{VHEGRB}). The simulation filters out alerts that don't meet observation requirements, including visibility and moonlight conditions. While moonlight observing capabilities differ across collaborations and evolved over time, we assume all three IACTs could observe under moderate moonlight from the start, using conservative conditions to preserve sensitivity. Between 2004 and May 2023, 1008, 913, and 886 alerts matched observation criteria for H.E.S.S., MAGIC, and VERITAS, respectively.

\begin{figure*}[!htb]
  \centering
\includegraphics[width=1\textwidth]{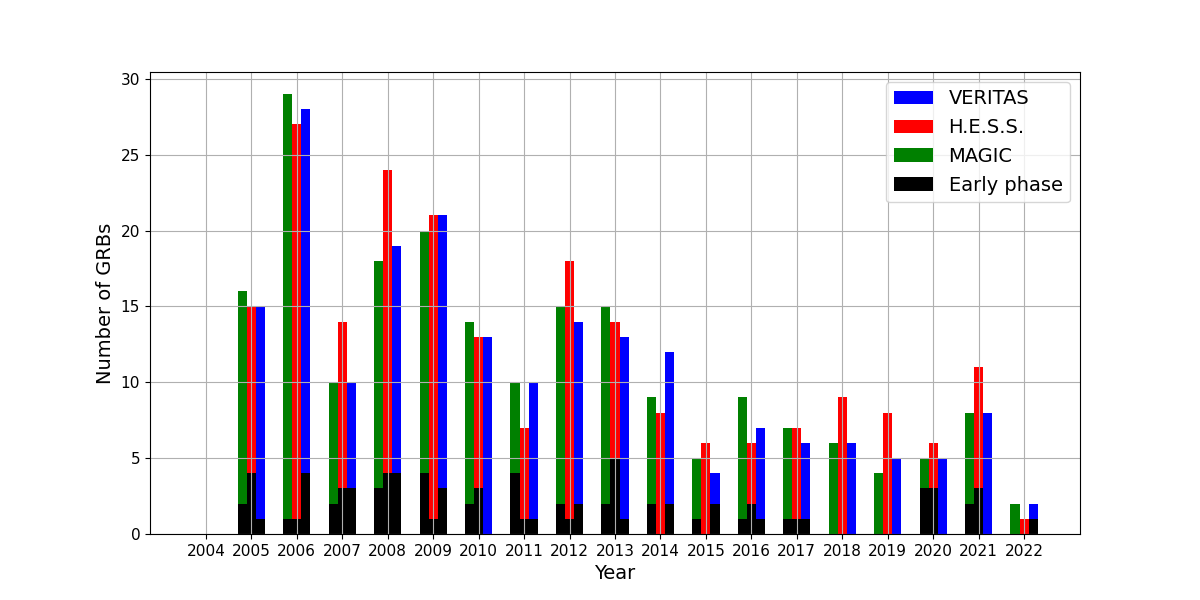}
\caption{Number of potential. GRB observation per year for H.E.S.S., MAGIC, and VERITAS. The GRBs that could have been observed with a delay of less than 600 seconds are marked in black as Early phase observations. We note that the MAGIC and VERITAS configurations considered in this work are only valid after 2007 and 2009 respectively and that H.E.S.S. only started implementing moonlight observations in 2019. From~\cite{VHEGRB}.}
\label{fig:GRB_OBS}
\end{figure*}

Valid alerts are filtered by confirming GRB classification and matching entries in the Swift GRB catalog~\cite{Swift_catalog}. Furthermore, we restrict our analysis to 488 GRBs with reliable redshift measurements and at least two consecutive Swift-XRT X-ray data points. From 2004 to June 2022, 215, 201, and 198 GRBs meet these criteria for H.E.S.S., MAGIC, and VERITAS, respectively—some observable by more than one IACT. We consider the full available observation window, though in practice, collaborations may observe for shorter durations. The number of potential observations per year is shown in Figure~\ref{fig:GRB_OBS}. 

\section{Analysis}
\label{sec:4}
\subsection{Step 1: Intrinsic Gamma-Ray Flux Estimation}

X-ray fluxes are obtained using the \texttt{swifttools} API\footnote{\url{https://www.swift.ac.uk/API}} to access \emph{Swift} Burst Analyser data~\cite{Evans_2010}. To ensure we are capturing the afterglow phase, we fit the X-ray light curves starting 2000 seconds after the GRB trigger, an empirical convention adopted for consistency. The intrinsic VHE gamma-ray flux is derived from the unabsorbed X-ray flux using assumption 1.

\subsection{Step 2: Gamma-Ray Flux on Earth (After EBL Absorption)}

Low-energy photon fields, particularly the EBL, absorb VHE gamma rays via pair production, reducing the observed flux on Earth compared to the intrinsic flux expected at a $\propto 1/D^2$ rate, where $D$ is the luminosity distance to the source. Other factors such as the GRB jet opening angle and the viewing angle also impact the observed flux. EBL absorption becomes significant beyond $z \sim 0.3$, and since most \emph{Swift} GRBs are located at $z > 1$, this attenuation is a primary reason for the lack of VHE GRB detections. We limit the analysis to GRBs with measured redshifts and use the Domínguez EBL model~\citep{2011MNRAS.410.2556D}. The attenuated flux observed on Earth $\mathrm{\phi^{(e)}}$ is then:

\begin{equation}
   \mathrm{\phi_{\gamma}^{(e)} = \phi_{\gamma}^{(u)} \, e^{-\tau(E, z)}}
\end{equation}

where $\mathrm{\tau(E, z)}$ is the energy- and redshift-dependent optical depth due to EBL absorption.

\subsection{Step 3: Gamma-Ray Flux as Seen by IACTs}

To estimate the flux observed by IACTs, we take into account their instrument-specific effective areas, which vary significantly with the zenith angle of observation. Publicly available effective area data are used for H.E.S.S., MAGIC, and VERITAS. The instrument energy range is defined between $\mathrm{E_1}$ and $\mathrm{E_2}$, and differs from the intrinsic energy bounds. The observed flux $\mathrm{\phi^{(obs)}}$ as seen by an IACT is calculated by integrating the attenuated flux over the instrument's energy range:

\begin{equation}
    \mathrm{\phi_{\gamma}^{(obs)} = \int_{E_1}^{E_2} \phi_{\gamma}^{(e)}(E) \, A_{\text{eff}}(E, \theta) \, dE}
\end{equation}

where $\mathrm{A_{\text{eff}}(E, \theta)}$ is the effective area as a function of energy $\mathrm{E}$ and zenith angle $\mathrm{\theta}$. Instruments with larger collecting areas (e.g., H.E.S.S. 28-m, MAGIC 17-m) allow for lower energy thresholds. 

\subsection{Step 4: Background Estimation}

In VHE gamma-ray astronomy, cosmic rays form a dominant source of background. Most analysis techniques employ filtering based on shower image parameters recorded by IACT cameras to reject non-gamma-ray events~\citep[e.g.,][]{1985ICRC....3..445H}.  Even after filtering a large amount of background events still exists. Background rates after event filtering, denoted as $\mathrm{R^B}$, are taken from publicly available sources for each IACT. The parameters used in steps 1 to 4 for each IACT are presented in~\cite{VHEGRB}.

\subsection{Step 5: Signal Significance Calculation}
To evaluate potential residual background contamination, we apply standard background estimation techniques~\citep{Berge_2007}. The analysis distinguishes between an ON region (target source) and one or more OFF regions (background), with event counts denoted as $\mathrm{N_{ON}}$ and $\mathrm{N_{OFF}}$, respectively. The relative exposure between these regions is expressed by the parameter $\mathrm{\alpha}$. The gamma-ray signal rate is given by: 

\begin{equation}
 \mathrm{   R(t) = R_0 \hspace{0.05cm} t ^{\alpha_{X}}}
\end{equation}

where the normalization $\mathrm{R_0}$ is computed by using assumptions 1 and 3.

The resulting VHE gamma-ray signal is computed as:

\begin{equation}
 \mathrm{   S = \int_{t_1}^{t_2} R(t) dt}
\end{equation}

The corresponding background contribution is:

\begin{equation}
  \mathrm{  B = \int_{t_1}^{t_2} R^B dt}
\end{equation}

with $\mathrm{t_1}$ and $\mathrm{t_2}$ denoting the start and end times of the observation. The expected counts in the ON region are given by $\mathrm{N_{ON} = S + \alpha B}$, while those in the OFF region are $\mathrm{N_{OFF} = B}$. The detection significance of the VHE gamma-ray signal is then computed following the method introduced by~\cite{1983ApJ...272..317L}.

\subsection{Step 6: Analysis of GRBs with High Significance}

\begin{figure*}
  \centering
  \begin{minipage}[b]{0.45\textwidth}
    \includegraphics[width=\textwidth]{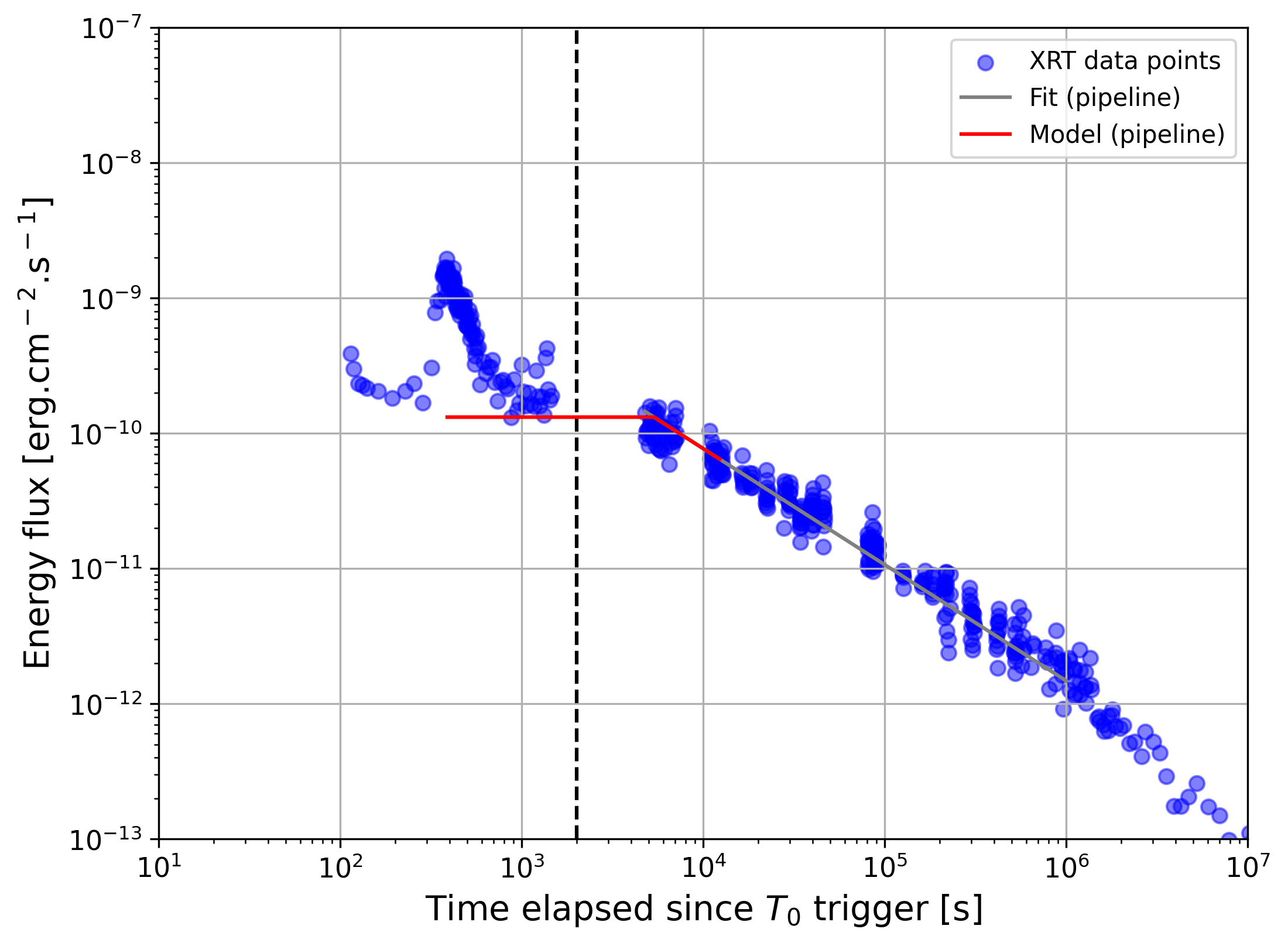}
  \end{minipage}
     \begin{minipage}[b]{0.45\textwidth}
    \includegraphics[width=\textwidth]{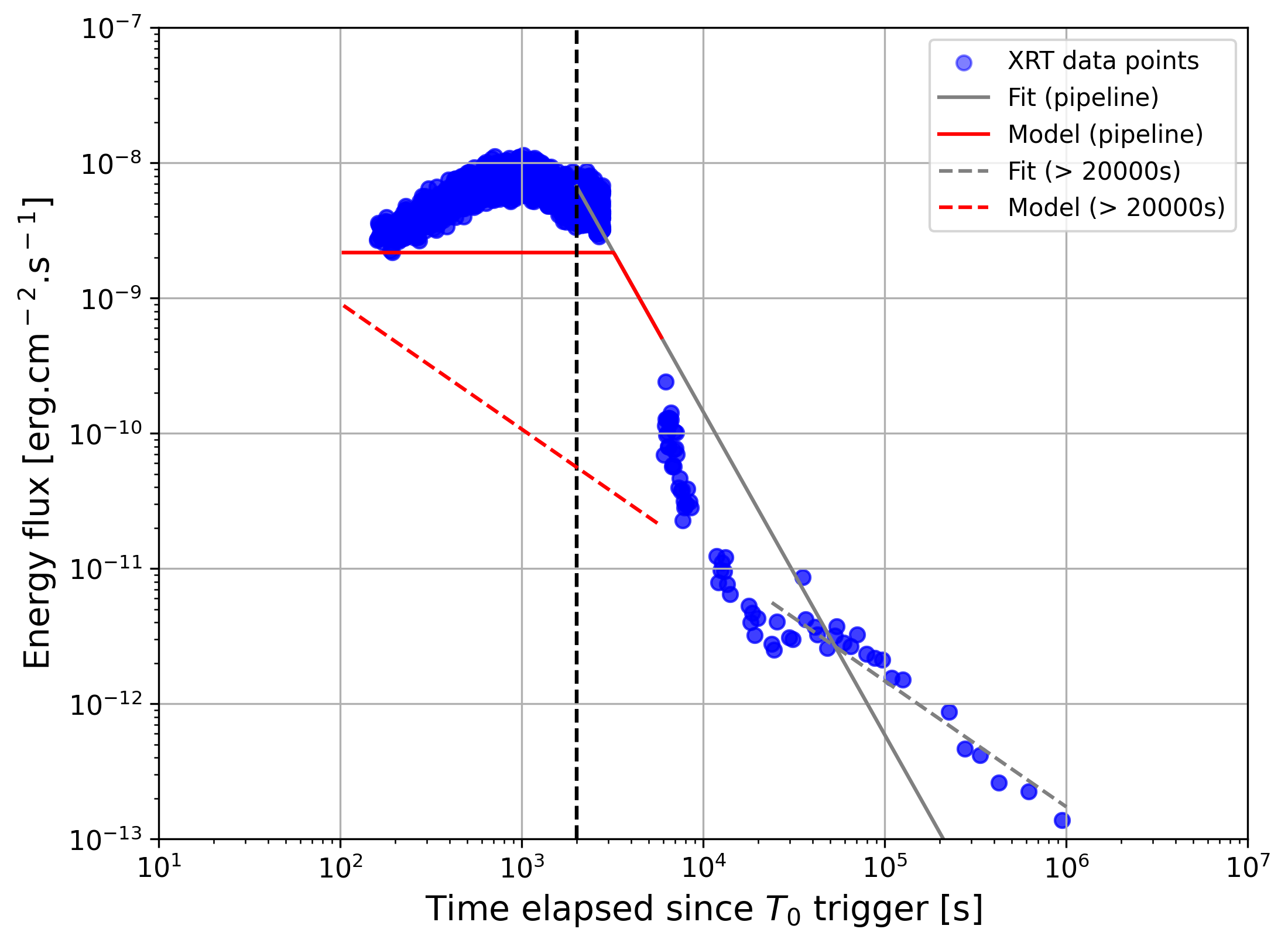}
  \end{minipage}
      \caption{X-ray light curve fits for GRB\,161219B (left) and GRB\,060218 (right). Blue dots represent \emph{Swift}-XRT data. In the pipeline, the minimum between the lowest X-ray point and the fitted curve is used to estimate the expected VHE signal during the H.E.S.S. (left) and VERITAS (right) observation windows, highlighted in red. The dotted gray and red lines on the right plot indicate a post-20000s fit, while the black dashed vertical line marks the 2000-second threshold. From~\cite{VHEGRB}.}
    \label{fig:xray_fit}
\end{figure*}

For GRBs exhibiting notable significance, we perform additional analysis of their X-ray light curves to more accurately constrain the simulated VHE signal. In particular, we verify that the early-time X-ray emission consistently exceeds the fitted model curve, as demonstrated for GRB\,060218 and GRB\,161219B in Fig.~\ref{fig:xray_fit}. Based on this, the computed significance over the full observational window can be interpreted as a conservative lower bound.

\section{Final Results}
\label{sec:5}
Among all the analyzed GRBs, the most promising cases for each instrument that could have been detected in the past are summarized in Table~\ref{tab:GRB_HESS}. For each GRB, we report the significance for a window starting after 2000 seconds and the computed detection significance over the full duration of the observation. A full list is given in~\cite{VHEGRB}.

\begin{table}[!htb]
\centering
\small 
\begin{tabular}{ccccccc}
\hline
\shortstack{GRB\\Name} & Instrument & z & \shortstack{Time\\(UTC)} & \shortstack{Delay\\(s)} & \shortstack{Dur.\\(s)} & \shortstack{$\sigma$ $>$ 2000s\\(full)} \\
\hline
GRB161219B & H.E.S.S. & 0.1475 & 2016-12-19T18:48:39 & 388 & 11899 & \textbf{11.5}  (\textbf{12.1}) \\
GRB101225A & MAGIC & 0.847  & 2010-12-25T18:37:45 & 1124 & 5622  & \textbf{4.8} (\textbf{5.4}) \\
GRB060218  & VERITAS & 0.0334 & 2006-02-18T03:34:30 & 104  & 5766  & \textbf{51.2}  (\textbf{56}) \\
\hline
\end{tabular}
\caption{GRBs potentially detectable by H.E.S.S., MAGIC and VERITAS. The table shows the GRB name, instrument, redshift, burst time, observation delay, observation duration, and detection significance for the windows after 2000 seconds and the full window.}
\label{tab:GRB_HESS}
\end{table}

\section{XZ table and prospects for CTAO}
\label{sec:6}
\begin{figure*}[!ht]
  \centering
\includegraphics[width=0.8 \textwidth]{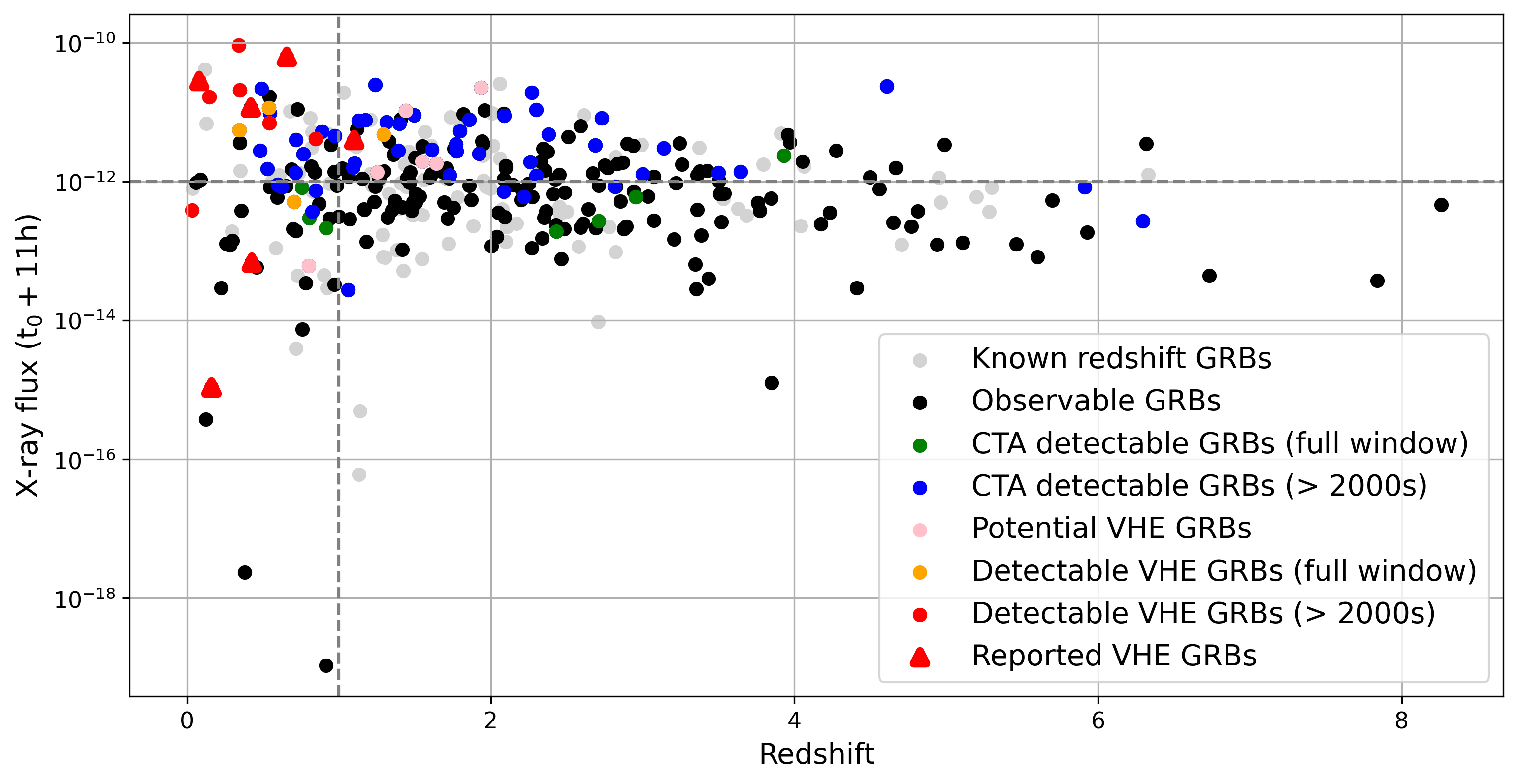}
\caption{GRB X-ray fluxes at 11 hours vs. redshift. The gray dots represent all GRBs that have a redshift measurement and an XRT X-ray flux at 11 hours (348 GRBs in total). The black dots represent the ones amongst them that are observable by the H.E.S.S., MAGIC, or VERITAS IACTS (248). The red triangles represent the GRBs for which VHE gamma-ray detections have been reported (6). The red dots represent in addition the GRBs that are detectable by current IACTs in the afterglow phase using data after 2000 seconds (8). The orange dots represent in addition the GRBs for which a prompt observation is possible and is detectable if data from the entire observation window is used (12). The pink dots represent in addition the GRBs that were initially flagged by our study as being potentially detectable at VHE energies (18). The blue and green dots are for CTAO the same as the orange and red dots respectively (65 and 71). From~\cite{VHEGRB}.}
\label{fig:XZplot}
\end{figure*}

We repeat our study assuming a hypothetical scenario where CTAO has been operational since 2004. CTAO, a next-generation IACT with greater sensitivity and lower energy thresholds, will operate from two sites: La Palma (north) and Chile (south)~\citep{2013APh....43..171B}\footnote{\url{https://www.cta-observatory.org/science/ctao-performance/}}. For simplicity, we adopt constant effective area and background rates across all zenith angles. This qualitative exercise is not a performance forecast but a comparison with current IACTs. We report our final findings in Figure~\ref{fig:XZplot} for all instruments.  CTAO's sensitivity to lower-energy gamma rays significantly improves detection prospects, especially in the second quadrant of the $\mathrm{XZ}$ plot, and increases the overall detection rate. However, results remain indicative due to site differences and simplified assumptions.

\section{Conclusion}
Our study shows that the detection rate of VHE GRBs remains very low under current-generation instruments: fewer than one per year globally, and under 0.5 per year for a single IACT site. With the upcoming CTAO Observatory, this rate could rise to approximately four detections per year accounting for GRBs with measured redshifts. Given the strict observational constraints, these rates are expected. We strongly encourage all IACT collaborations to revisit archival data for the GRBs highlighted in this work, as previously unexamined signals may be present.

Future efforts should aim at relaxing observation criteria and improving response times. The non-detection of GRB\,221009A~\cite{GRB22109A}, despite its strong X-ray signal even 51 hours post-burst, suggests that some of our assumptions—such as the flux normalization—may need to be re-evaluated at later times after the burst. Ultimately, a deeper understanding of the link between X-ray and VHE emission at early and late times will require joint analyses, such as stacked searches, to constrain emission mechanisms and improve detection strategies.

\end{document}